\documentclass{elsart3p}

\usepackage{graphicx} 
\usepackage{epstopdf} 
\usepackage{ifpdf}
\usepackage{graphicx,amssymb,lineno}
\ifpdf
\usepackage[%
  pdftitle={Instructions for use of the document class
    elsart},%
  pdfauthor={Simon Pepping},%
  pdfsubject={The preprint document class elsart},%
  pdfkeywords={instructions for use, elsart, document class},%
  pdfstartview=FitH,%
  bookmarks=true,%
  bookmarksopen=true,%
  breaklinks=true,%
  colorlinks=true,%
  linkcolor=blue,anchorcolor=blue,%
  citecolor=blue,filecolor=blue,%
  menucolor=blue,pagecolor=blue,%
  urlcolor=blue]{hyperref}
\else
\usepackage[%
  breaklinks=true,%
  colorlinks=true,%
  linkcolor=blue,anchorcolor=blue,%
  citecolor=blue,filecolor=blue,%
  menucolor=blue,pagecolor=blue,%
  urlcolor=blue]{hyperref}
\fi

\makeatletter
\def\elsartstyle{%
    \def\normalsize{\@setfontsize\normalsize\@xiipt{14.5}}
    \def\small{\@setfontsize\small\@xipt{13.6}}
    \let\footnotesize=\small
    \def\large{\@setfontsize\large\@xivpt{18}}
    \def\Large{\@setfontsize\Large\@xviipt{22}}
    \skip\@mpfootins = 18\p@ \@plus 2\p@
    \normalsize
}
\@ifundefined{square}{}{}
\makeatother

\begin{document}

\begin{frontmatter}
\title{Approaches to Single Photon Detection}
\author{R.~T.~Thew$^{1}$, N.~Curtz$^{1,2}$, P.~Eraerds$^{1}$, N.~Walenta$^{1}$,  J-D.~Gautier$^{1}$, E.~Koller$^{2}$, J.~Zhang$^{1}$,}
\author{N.~Gisin$^{1}$ and H.~Zbinden$^{1}$}
\author{}
\address{$^{1}$Group of Applied Physics, University of Geneva, 1211 Geneva 4, Switzerland}
\address{$^{2}$Department of Condensed Matter Physics, University of Geneva, 1211 Geneva 4, Switzerland}
\ead{robert.thew@physics.unige.ch}
\ead[url]{www.gapopotic.unige.ch}

\begin{abstract}
We present recent results on our development of single photon detectors, including: gated and free-running InGaAs/InP avalanche photodiodes; hybrid detection systems based on sum-frequency generation and Si APDs; and SSPDs (superconducting single photon detectors), for telecom wavelengths; as well as SiPM (Silicon photomultiplier) detectors operating in the visible regime.
\end{abstract}

\end{frontmatter}

\section{Introduction}
\label{intro}
Single photon detectors provide the ultimate in sensitivity for low light level systems. As more and more systems approach the regime of a few photons, or indeed, single photons, the demands on these detectors becomes even higher. However, the sensitivity, or  detection efficiency, is but one of many parameters that needs to be considered depending on the final application. In this regard, our approach is two fold: on the one hand, focusing on the characterisation and development of the detectors themselves: and on the other,  adapting them and developing the interfaces necessary for their utilisation in applications and metrology. 

The emergent field of quantum communication \cite{GisinThew}, which includes the newly commercialised technology of quantum key distribution  \cite{Gisin02,qkdCom} and quantum random number generation \cite{Stefanov99}, are typical examples of the applications and technology currently motivating us. However, we have also shown their suitability for metrology in distributed measurement schemes such as optical time domain reflectometry (OTDR) \cite{Legre07}. More fundamentally, there has  also recently been a push towards their role in radiometric scales for the redefinition of the Candela in terms of photon number \cite{Cheung07}. Detectors are an enabling technology and as the quality and diversity of single photon detectors improves and increases we expect to see their impact in more and more diverse fields of research, both fundamental and applied.

We are working with a variety of different approaches for single photon detection and here we give a brief overview of the capabilities of several of these state-of-the-art single photon detectors and elaborate on their positive and negative traits. Their general characteristics are summarised in Table \ref{Tab:Comparison}.


\section{InGaAs/InP APDs}
InGaAs/InP APDs have been the workhorse single photon detector in the telecommunication wavelength regime for the last ten years. Whilst robust, they have suffered from relatively high noise levels and after-pulsing  \cite{Ingaas}. This has restricted their operation to the gated regime, i.e. turned on when photons are expected, and required the application of so called dead-times, where the detector is  turned off, usually for around 10$\mu$s, after a detection event. Steady improvement in the diode characteristics and the operating electronics has  seen new regimes open up, and with that, new possibilities for applications.
\begin{figure}[!t]
\begin{center}
\includegraphics*[width=7cm]{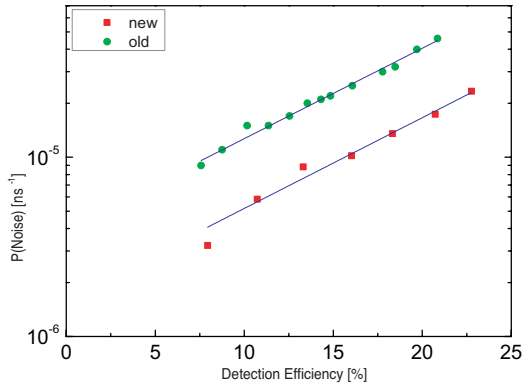}
\caption{\label{fig:Ingaas1} Comparison of efficiency and noise characteristics for an APD with conventional electronics (old) and the active quenching ASIC (new). Significant advantages in both noise and after-pulsing (not shown) have been achieved.}
\end{center}
\end{figure}

\subsection{Gated Regime}
The "standard" regime of operation for InGaAs/InP APDs \cite{Ingaas} is to gate them when a photon is expected. This has been used to great effect in QKD but also trains of gates can be used,  for example,  to improve the measurement times and statistics in OTDR. Recently, we have developed an application specific integrated circuit (ASIC) to control the gating and quenching operation of these type of APDs. Typical detection efficiency and noise, dark count probability per ns (gate width), are shown in Fig. \ref{fig:Ingaas1} and we clearly see a significant decrease in the noise, for a given efficiency, that is due to this circuit \cite{Thew07}. The key to the noise reduction in the gated mode is the rapid quenching, turning off, of the APD after a detection. Recent optical characterisation has shown that the quenching time is of the order of $\sim$ 100\,ps \cite{Zhang08}.  This ability to rapidly close the gate also has the flow on effect of reduced after-pulsing.

\subsection{Free Running and "Long" Gates}
\begin{figure}[!t]
\begin{center}
\includegraphics*[width=8cm]{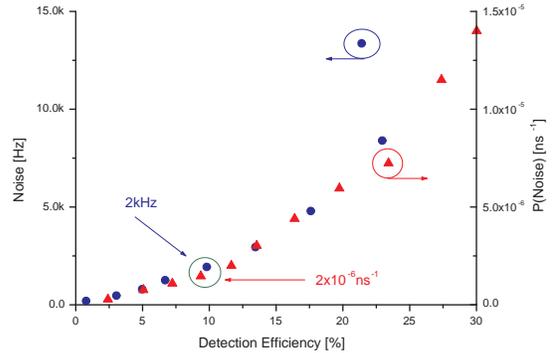}
\caption{\label{fig:Ingaas2} Comparison of efficiency and noise characteristics in gated and free-running modes for the same APD showing no loss of performance between the two regimes.}
\end{center}
\end{figure}

The inherently compact nature of the ASIC significantly reduces parasitic capacitance, which permits the gates to be held on over much longer times. Normally  this parasitic capacitance corresponds to a voltage leakage and subsequent reduction in detection efficiency until the APD is no longer sufficiently biased. The resolution of this problem opens up the possibility for the first time of not only extending the gate length, from a few ns to $O$($\mu$s), but further still, to a free running regime. The detection efficiency and noise for the two modes of operation (in Hz for free-running or per gate) are in good agreement as we  see in Fig. \ref{fig:Ingaas2}. More details of the initial scheme can be found elsewhere \cite{Thew07} and a more detailed analysis is currently in preparation \cite{Zhang08}. 


\section{Superconducting Single Photon Detectors}
Superconducting Single Photon Detectors are a relatively new approach for low light detection \cite{Goltsman01}. Despite the need for cryogenic temperatures they hold great promise for overall performance in terms of detection efficiency, noise, timing jitter and maximum count rates. There are also initial results for photon number resolution \cite{Divochiy08}.

These detector's operation is dependent on the absorption of a photon on a superconducting meander (wire) where the subsequent heating forces a superconducting-normal transition which is then read-out. These meanders are made out of NbN and are around 100\,nm wide and a few nm thick. The overall detection area, not including the fill factor, is around 10\,$\mu$m$^2$. The detectors are normally cooled in liquid Helium (4.2\,K). In this configuration devices have been fabricated with QE $\sim$ 10\% (at 1300\,nm). The need for cryogenic cooling is certainly a barrier to the practical operation of these devices but this is off-set by the detection performance. Recently we have been working on these devices operating at $\sim$\,3\,K and adapted for operation in closed cycle coolers,  providing a major step towards their practical implementation \cite{Walenta08}. 

\begin{figure}[!t]
\begin{center}
\includegraphics*[width=7.5cm]{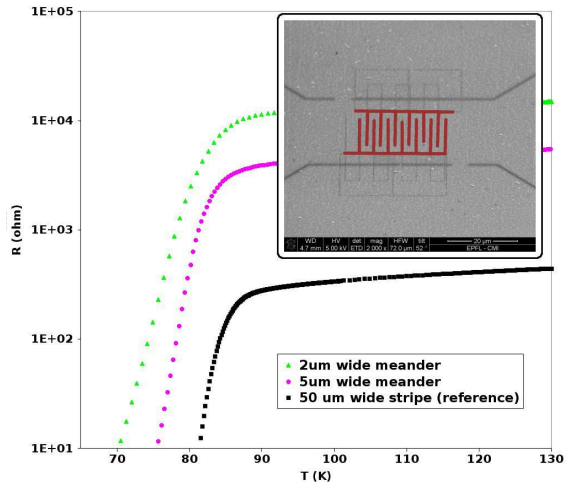}
\caption{\label{fig:sspd} The Resistance vs Temperature behaviour for the first High-T$_c$ SSPD samples and inset a micrograph of the 2$\mu$m-wide meander sample.}
\end{center}
\end{figure}
To increase the practicality of these devices, we are currently looking at the long term possibility of using different materials, such as YBa$_2$Cu$_3$O$_{7-\delta}$, for High-T$_c$ superconducting detectors, with the perspective to operate with liquid Nitrogen (77\,K) instead of Helium. Along with this new material comes the need to develop new fabrication processes, but already significant progress has been made. In Fig. \ref{fig:sspd} we see the electrical response for samples with meanders at 5\,$\mu$m and 2\,$\mu$m. However, this is still some way from the 100\,nm wide structures currently achieved in NbN. The next stage of development will see the optical characterisation of these devices. Once in this temperature regime the demands on cooling, especially in terms of cost and complexity, should be considerable reduced.


\section{SFG-Si APDs}
A possible compromise in terms of efficiency, noise, jitter and complexity is the up-conversion, or sum-frequency generation, detectors. This detection scheme uses a nonlinear optical process to convert the photons that one wishes to study to a regime where Silicon (Si) APDs have good characteristics. This is a scheme that we have used quite successfully for telecom wavelengths in QKD \cite{Thew06} as well as for showing an order of magnitude improvement in OTDR measurement precision \cite{Legre07}.  We have predominantly used periodically poled Lithium niobate (PPLN) waveguides for the nonlinear conversion process at telecom wavelengths \cite{Thew06}. However, we are not restricted to this regime and there are a range of possible nonlinear crystals (NLC) for the up-conversion process depending on the desired interaction. We have previously shown this system working for single photon detection in the mid-infrared at 4.6\,$\mu$m \cite{Karstad05}.

\begin{figure}[!t]
\begin{center}
\includegraphics*[width=8cm]{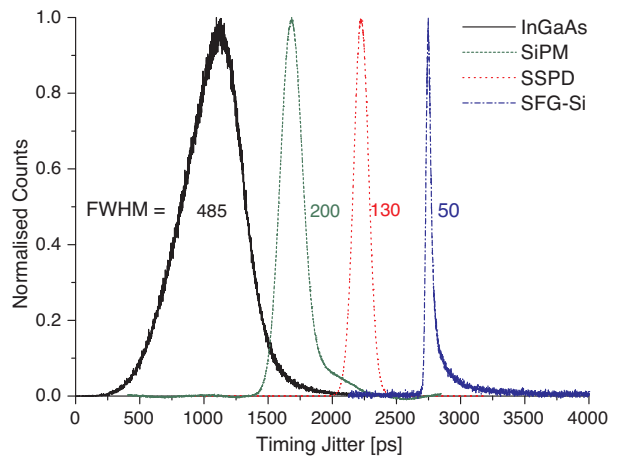}
\caption{\label{fig:Jitter} Comparison of timing jitter for the InGaAs, SSPD, SFG-Si and SiPM detectors with FWHM given in ps.}
\end{center}
\end{figure}
Apart from the ability to adopt Silicon APDs for single photon detection at a wide variety of previously unobtainable wavelengths, the timing resolution of some of the new generation of Si APDs is exceptionally good. We see in Fig. \ref{fig:Jitter}, that the 50\,ps jitter compares favorably with the SSPD detectors (other SSPDs have shown jitter down to 20\,ps) and is significantly lower than for the InGaAs/InP APDs. Overall QEs of more than 10\,\% are possible although there are currently unresolved (nonlinear) noise issues that force them to be used at lower efficiencies as indicated in Table \ref{Tab:Comparison}.


\begin{table*}[t!]
\begin{center}
\caption{Comparison of the key detector parameters for devices currently under test and development: QE = Quantum Efficiency; Prob. (A-P) = After-pulse probability; $^{*}$  depends on the NLC; $^{\dag}$  this is for the device under test, other SSPDs have been reported with jitter as low as 20\,ps.) }\label{Tab:Comparison} \vspace{2mm}
\begin{tabular}{|l|c|c|c|c|c|c|c|}\hline\hline
Detectors  & QE [\%] & Noise & Prob. (A-P) & Jitter [ps] & Max. Count [MHz] & Temp. [K] & $\lambda$ [nm]  \\ \hline  \hline
InGaAs (Gated) & 10 (20) & 2x10$^{-6}$\,ns$^{-1}$  (6x10$^{-6}$\,ns$^{-1}$)& $<10^{-3}$ns$^{-1}$  &  485  & 0.1 (lim. dead time) & 223 & 1000-1600  \\ \hline
InGaAs (Free) & 10 (20) & 2\,kHz (6\,kHz) & $<$ 3\% & 485   & 0.1 (lim. dead time) & 223 &   1000-1600    \\ \hline
SSPD &  2.5 (10) & 10\,Hz ($>$10\,kHz) &  - & 130$^{\dag} $ & 50 (potentially $\>$\,1000) & 2 - 4.2  &  VIS-NIR \\ \hline
SFG-Si & 2 (12)   &  2\,kHz (250\,kHz) &  $<$ 1\% & 50  & 20  & NLC @ 350 & 400-5000$^{*}$ \\ \hline
SiPM &    16 &  52\,kHz &  - & 200   & 430  &  268 &  400 - 1000\\ \hline
\end{tabular} 
\end{center}
\end{table*}

\section{SiPM}
The final detector that we are looking at is a CMOS fabricated array of Si APDs \cite{Eraerds07}. These devices, called Silicon Photo-multipliers (SiPMs), have been commercially available for several years now. They were originally developed to replace classical photomultipliers in nuclear applications though more recently they are finding use in bio-medical systems. Typically, they consist of 100-1000 parallel connected photodiodes mounted on a common chip.  We are using a SiPM (id Quantique) with 132 photodiodes with a surface area of 780\,$\mu$m\,x\,780\,$\mu$m of which 31\,\% (fill factor) is photosensitive. We operate the device in the Geiger mode. As is normal in this mode, if there is an avalanche in one of the photodiodes, there is an electrical pulse at the output of the SiPM. One of the useful characteristics of these SiPMs is that if two or more avalanches occur at the same time, the amplitude of the electrical output pulse is equal to the number of avalanches times the amplitude of a single avalanche. Therefore, the height of the output pulse is also proportional to the number of detected photons, providing photon number resolving capabilities.

We are currently not so much interested in the photon number resolving characteristics of these SiPMs, primarily due to the relatively low QE. However, what these detectors do provide is a system for very high count rates. We have currently measured peak detection rates of around 430\,MHz. This lends itself to integration in novel schemes for fast counting, such as in  quantum random number generators \cite{Stefanov99}, where current commercial devices are limited to around 4\,MHz and as such there is room for significant improvements.


\section{Conclusion}
We have summarised our efforts towards the test and development of state of the art single photon detectors. Our motivation for this work, though not discussed here, is predominantly quantum communication though this also reinforces our work in telecommunication metrology. The demands of quantum communication are providing the impetus for rapid improvements in detection technologies and this is having a flow on effect into many emerging fields of research where low light levels need to be detected. We expect that the range of detectors we are developing  will have a significant impact  and provide solutions for low light level optical metrology. 

\vspace{5mm}
The authors  acknowledge financial support from the European projects SINPHONIA and QAP and
Swiss NCCRs, Quantum Photonics and MANEP.

\end{document}